\documentclass[showpacs,preprintnumbers,prl,twocolumn]{revtex4}
\usepackage{amssymb}
\usepackage{amsmath}
\usepackage{graphicx}
\usepackage{bm}
\usepackage{amsfonts}

\def\be{\begin{equation}} 
\def\ee{\end{equation}}

\begin{document}

\title{Readout for Phase Qubits without Josephson Junctions}

\author{Matthias Steffen}
\email{msteffe@us.ibm.com}
\author{Shwetank Kumar}
\author{David DiVincenzo}
\author{George Keefe}
\author{Mark Ketchen} 
\author{Mary Beth Rothwell}
\author{Jim Rozen}

\affiliation{IBM Watson Research Center, Yorktown Heights, NY 10598}



\begin{abstract}
We present a novel readout scheme for phase qubits which eliminates the read-out
SQUID so that the entire qubit and measurement circuitry only requires a single
Josephson junction. Our scheme capacitively couples the phase qubit directly to
a transmission line and detects its state after the measurement pulse by
determining a frequency shift observable in the forward scattering parameter of
the readout microwaves. This readout is extendable to multiple phase qubits
coupled to a common readout line and can in principle be used for other flux
biased qubits having two quasi-stable readout configurations.
\end{abstract}

\volumeyear{year}
\volumenumber{number} 
\issuenumber{number}
\eid{identifier}
\date{\today}

\maketitle

The reliable identification of the quantum state occupation probability of
a qubit with high fidelity without introducing detrimental system complexity
has been at the center of attention in the design and operation of
superconducting qubits in recent
years\cite{Nakamura99,Martinis02,Lang03,Robertson05}. Over the years these
trade-offs were optimized for various types of qubits to enhance overall qubit
performance \cite{Wallraff04,Lupascu04,Siddiqi06}. Although some of these new
methods could also apply to the phase qubit, measurement techniques for these
qubits have remained largely unchanged. The switching of a two or three
Josephson junction SQUID into its normal state remains the main technique to
identify the qubit state of a phase qubit
\cite{Palomaki06,Ansmann09,Hofheinz09}.

In this letter, we show a novel phase qubit read-out method \cite{Steffen08}
which eliminates the SQUID and its associated dissipation entirely \cite{Lang03}. By removing the
SQUID we also reduce the number of required Josephson junctions from three or
four to only a {\em single} Josephson junction, thereby dramatically improving
the overall device yield.

The measurement of phase qubits is typically done in two steps (e.g.
\cite{Cooper04}). First, the quantum state is "measured" by taking advantage of
the potential energy landscape such that post measurement the system is in one
of two quasi-stable configuration depending on the quantum state prior to
measurement. This step has been demonstrated with large single-shot fidelities
and reliable performance \cite{Ansmann09}. Finally, the qubit is "read out" by
identifying which of the quasi-stable configurations the system is in. By
convention we refer to these configurations as the left (L) and right (R)
configuration. Because the L and R configurations correspond to a large
difference of the circulating current in the qubit, the obvious choice has been
to flux couple a SQUID to the qubit and detect the current at which a SQUID
switches to its voltage state. The configuration, L or R, can be identified with
high confidence using the SQUID, usually without significant backaction.
Recently, however there has been mounting evidence that the SQUID does indeed
limit qubit coherence times or at least drastically impacts the repetition rate
of the experiment \cite{Lang03,commentf}. 

\begin{figure}[h]
\includegraphics[width=0.5\textwidth]{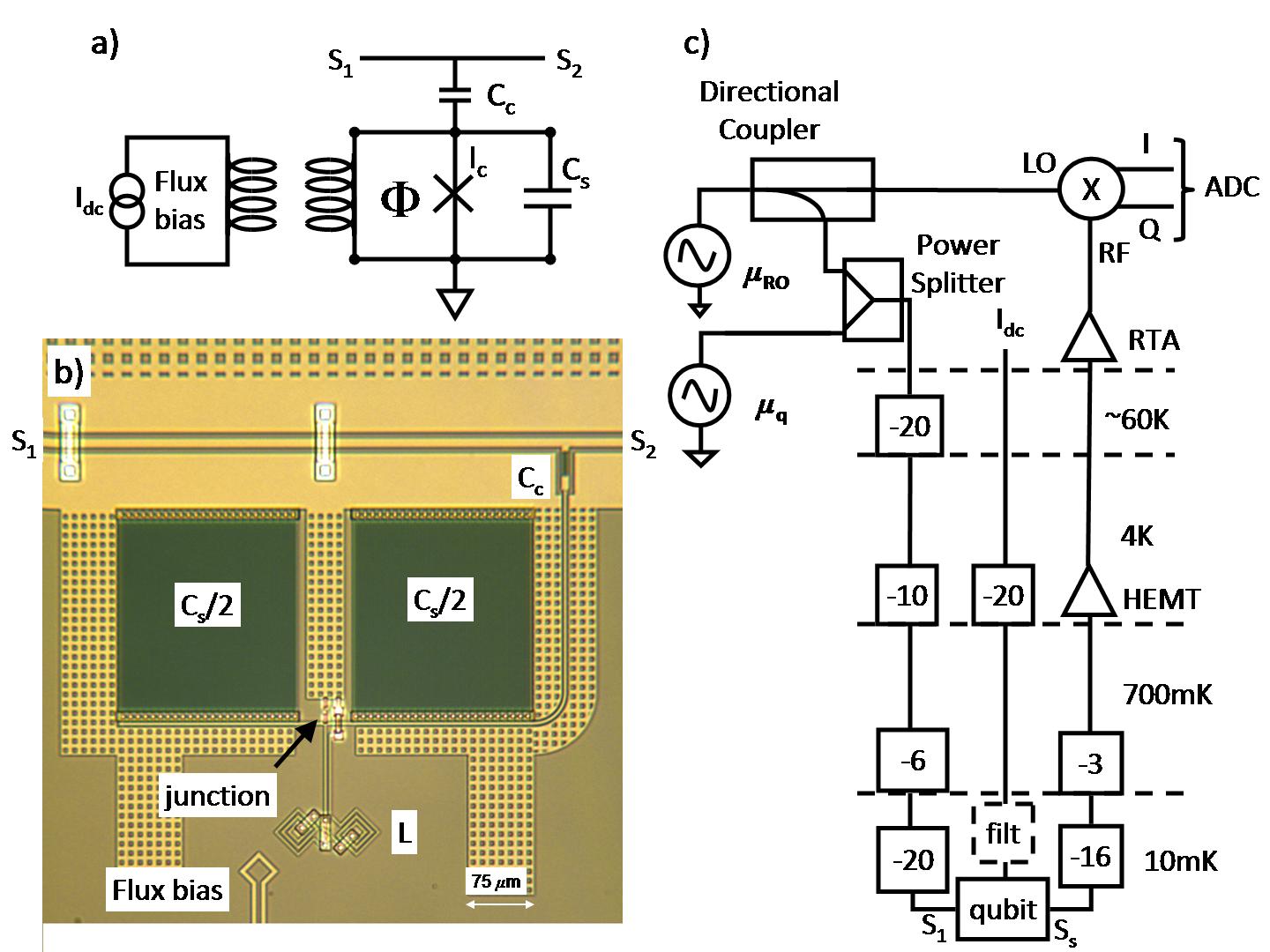}
	\caption{Outline (a) and micrograph (b) of the qubit. The phase qubit
consists of the standard circuit elements in parallel - a Josephson junction, an
inductor and a capacitor all capacitively coupled to a feedline via a coupling
capacitor. An optical micrograph of the qubit depicts the actual layout of
circuit. c) Shows the readout schematic, The qubit flux bias and microwave lines
(S1,S2) are sufficiently filtered and attenuated to ensure sufficiently low
electron temperatures. A HEMT at the 4K stage amplifies the outgoing microwave
signal.}
		\label{fig:fig1}
\end{figure}

Because the L and R configurations in a phase qubit have dramatically different
resonance frequencies we propose to identify the system configuration by probing
the qubit with microwave pulses using dispersive techniques
\cite{Wallraff04,Lupascu04,Siddiqi06}. We estimate that the correct
identification of the L or R configuration is possible with high certainty given
read-out times comparable to those using the SQUID. The new technique is made
possible by directly coupling the qubit to a microwave feedline and probing the
amplitude or phase response of a microwave pulse that passes by the qubit.
The new readout is compatible with all existing state-of-the-art phase qubit
techniques, including reset and measurement.

We have successfully implemented the most basic operations of this microwave
based read out technique. Here we present our experimental results and speculate on
further improvements. Because the operation and
"measurement" of the phase qubit does not change dramatically in our
implementation we are able to base our design on published literature
\cite{Cooper04,Ansmann09,Hofheinz09}. The basic layout of the qubit circuit is
shown in Fig. \ref{fig:fig1}a. We chose a capacitively shunted phase qubit to
minimize the number of junction two-level systems in the qubit spectroscopy and
maximize measurement fidelity \cite{Steffen06a}. In order to ensure long
coherence times of the qubit we fabricated the shunting capacitor using an
interdigitated comb with the ends shorted together to eliminate parasitic cavity
modes \cite{commentf}. The target value for the shunting capacitor is $C_s=2$
pF.
The qubit loop is designed gradiometrically with $L=625$ pH to eliminate
sensitivities to stray flux variations and reducing flux coupling to the
microwave feedline. The coupling capacitor is chosen to be $C_c=23$ fF which
should limit $T_1$ to about $T_{1,feed}=2C/(Z_0(\omega C_C)^2) \approx 180$ ns
for a qubit frequencies near $4.6$ GHz \cite{Oconnell08}.

\begin{figure}[h]
		\includegraphics[width=0.5\textwidth]{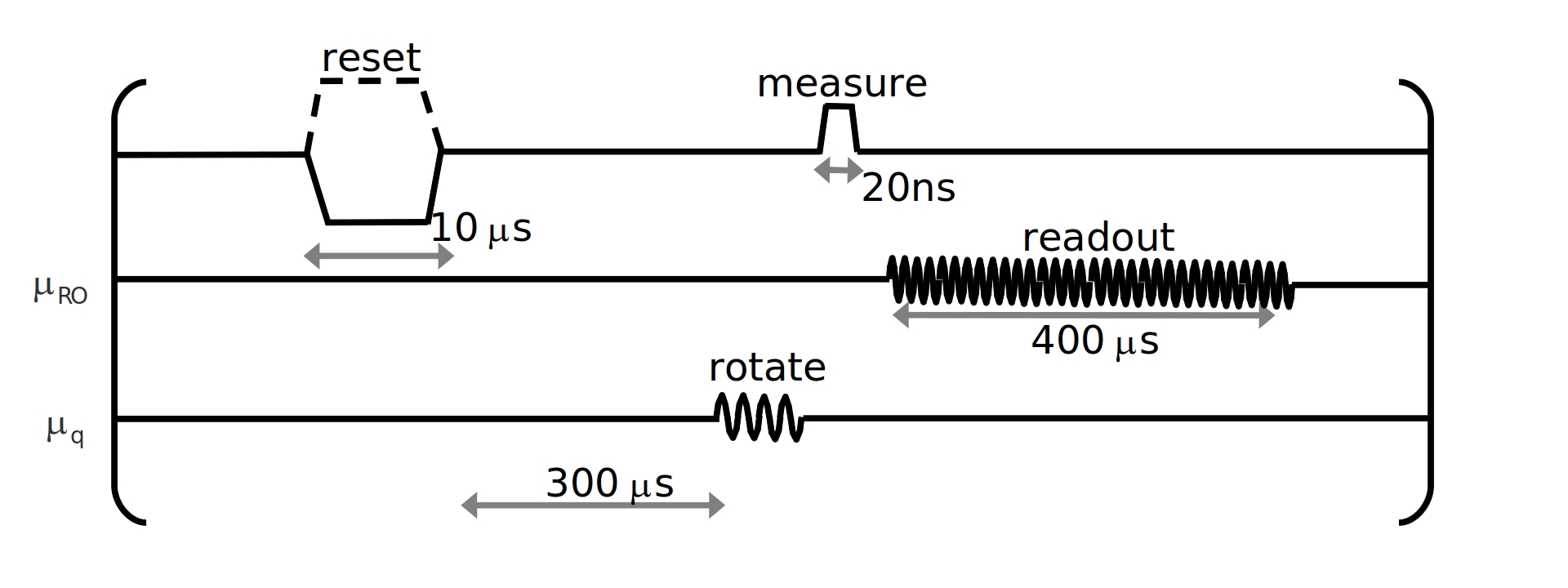}
	\caption{Experimental pulse sequence. The flux bias $\Phi$ and qubit
microwave pulse $\mu_q$ follow standard phase qubit protocols. The microwave
read-out pulse $\mu_{RO}$ is applied after the measurement pulse is executed for
a duration of $400\mu s$.}
		\label{fig:fig2}
\end{figure}

The fabrication was similar to ref \cite{Steffen10}. An optical micrograph
of the fabricated qubit is shown in Fig. \ref{fig:fig1}b. The sample was mounted
inside an RF-tight box and cooled down in a dilution refrigerator. The bias
lines were configured as outlined in Fig. \ref{fig:fig1}c. The flux bias line is
filtered with a low pass bronze powder filter, matched to $Z_0=50$ $\Omega$
impedance, with a $1$ GHz cut-off frequency \cite{Milliken07}.

\begin{figure}[h]
		\includegraphics[width=0.5\textwidth]{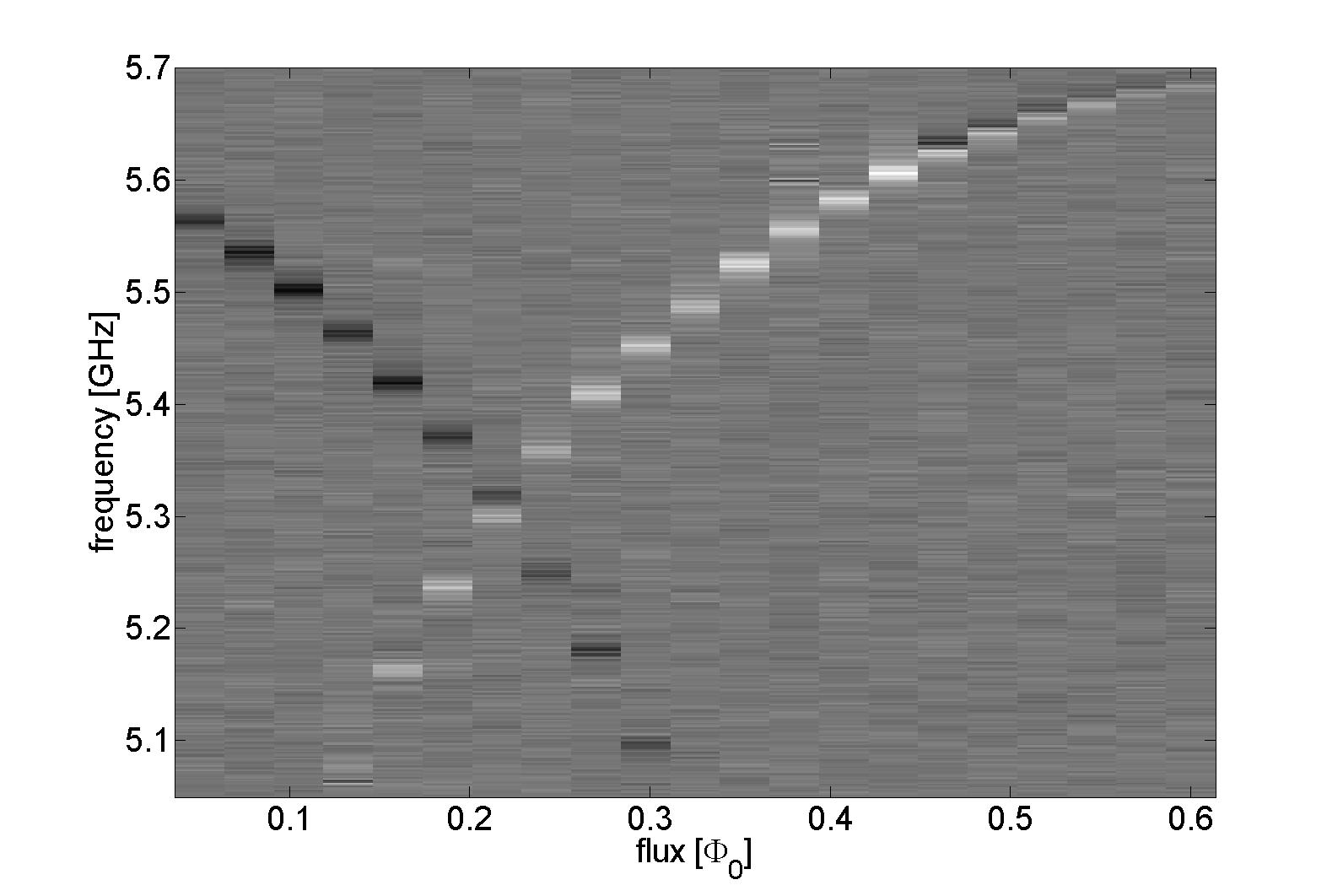}
	\caption{Frequency response of the phase qubit for negative (black) and
positive (white) reset pulses. When the qubit is hysteretic two resonance
frequencies are observed, consistent with the L and R configurations. The lower
frequency response $\omega_L$ corresponds to the configuration in which the
qubit is operated, and the other, $\omega_R$, corresponds to the configuration
in which the phase particle tunneled during the measurement pulse. The read-out
is performed by observing the presence or absence of resonance response at
$\omega_R$.}
		\label{fig:fig3}
\end{figure}

The qubit is calibrated by executing a pulse sequence similar to the one shown
in Fig. \ref{fig:fig2}. A reset pulse is applied to the flux line and added to
the flux bias line via a DC-block at room temperature. However, neither the
microwave pulse "rotate" nor the subsequent measurement pulse "measure" is
applied to the qubit. The microwave readout signal $\mu_{RO}$ is mixed with
itself rather than a separate microwave source tuned to the same frequency to
eliminate phase drift between the microwave generators. The resultant I and Q
signals at DC are filtered and digitized using an Acqiris data acquisition board
and averaged on board using sufficient averages for acceptable signal-to-noise
ratios. We then plot the amplitude response $|I+iQ|$ versus flux in Fig.
\ref{fig:fig3}. To better visualize the results we plot the negative amplitude
for a negative reset pulse (black) and the positive amplitude for a positive
reset pulse (white). Whenever the qubit is hysteretic (between $0$ and $\approx
0.45 \Phi_0$) the frequency response is vastly different depending on which
reset pulse was applied, showing that the L and R configurations have very
different resonant frequencies.

We now set the DC flux to $0.425 \Phi_0$ which is close to the edge of where
hysteresis observed. Depending on the reset pulse two resonance frequencies can
be observed. The lower one corresponds to the configuration in which the phase
is located in the shallow well (L configuration) - this is the configuration
where the qubit is operated. The higher one, $\omega_R$, corresponds to the R
configuration. We are interested in $\omega_R$ because this is the expected
resonant frequency after the phase tunnels from the L to R configuration during
measurement. From here on all experiments follow standard phase qubit protocols
with the only difference being how the signal is measured. In this case we
measure the amplitude response at $\omega_R$ with respect to the reference,
defined by the result obtained when the measurement pulse is absent. In
principle we only need to calibrate the reference once, but we obtain improved
results by continually recording the reference value because of small drifts.
Also note since we choose to average the homodyne voltage on the data
acquisition card, we are not performing a single shot read out. Due to the $19$
dB of attenuation the signal is reduced almost a factor of ten so that the
high fidelity single shot acquisition times for $\mu_{RO}$ would have been
prohibitively long ($10$ ms assuming the experiment is limited to the noise
generated by the HEMT). By replacing the attenuators with properly designed
isolators it will be possible to reduce the required acquisition times to
$100-500$ $\mu$s, comparable to those using a SQUID based single-shot read out.

\begin{figure}[h]
	\begin{eqnarray}
		\includegraphics[width=0.5\textwidth]{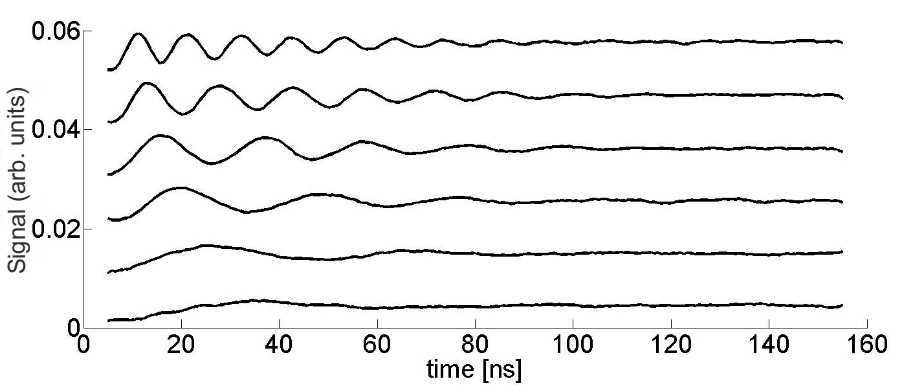} \nonumber \\
		\includegraphics[width=0.5\textwidth]{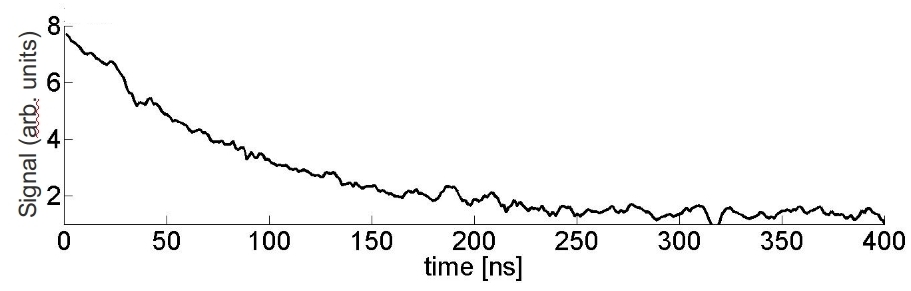} \nonumber
		\end{eqnarray}
	\caption{Rabi oscillations and an energy relaxation curve of the qubit.
Note that no absolute scale is shown as we are presently not performing a
single shot measurement.}
	\label{fig:fig4}
\end{figure}

Following standard phase qubit pulse sequences we are able to characterize the
qubit. We show in Fig. \ref{fig:fig4} the results for Rabi oscillations and a
$T_1$ experiment for a qubit frequency of
$\omega_L/2 \pi=4.46$ GHz ($\omega_R/2 \pi=5.6$ GHz). The energy relaxation time $T_1$ is found to be approximately $83$
ns. We believe this value is limited by $T_{1,feed}$ because a separate sample
using a SQUID based read out gave $T_{1,instrinsic}=260$ ns and because
$(1/T_{1,intrinsic}+1/T_{1,feed})^{-1}=106$ ns is close to the observed
coherence time. An improvement in $T_1$ should therefore be possible by simply
reducing the size of the coupling capacitor.

We have clearly demonstrated the basic principle of a new microwave method of
reading out phase qubits. It is instructive to discuss how the current
implementation relates to other experimental practicalities as well as
to scalability. The most significant drawback of our technique is the fact that
the $T_1$ times of the qubit are somewhat affected by the coupling to the
feedline. This is perfectly acceptable, however, if one is interested in
studying qubit spectroscopy \cite{Kline09}. However, because of the large size
of frequency shifts for the L and R configurations we believe that it should be
possible to also circumvent this limitation. By coupling the qubit to other
linear elements we estimate it should still be possible to
observe a frequency shift dependent on the qubit configuration without limiting
coherence times \cite{commentf}. We note that our proposed scheme is a step towards multiplexing many phase qubits off a single feedline, similar to readout schemes for photon detectors \cite{Day03}. 

\bibliography{bibmaster}

\end{document}